\documentstyle[floats,prl,aps]{revtex}
\begin{document}
\renewcommand{\thefootnote}{\fnsymbol{footnote}}
\draft
\title{\large\bf 
Eight state supersymmetric $U$ model of strongly correlated fermions} 

\author{Mark D. Gould, Yao-Zhong Zhang \footnote {Queen Elizabeth II Fellow.
                                   E-mail: yzz@maths.uq.edu.au}
             and 
        Huan-Qiang Zhou \footnote {On leave of absence from Dept of
	         Physics, Chongqing University, Chongqing 630044, China.
                 E-mail: hqzhou@cqu.edu.cn}} 

\address{      Department of Mathematics, University of Queensland,
		     Brisbane, Qld 4072, Australia}

\maketitle

\vspace{10pt}

\begin{abstract}
An integrable eight state supersymmtric $U$ model is proposed, which is a
fermion model with correlated single-particle and pair hoppings as well as
uncorrelated triple-particle hopping. It has an $gl(3|1)$ supersymmetry
and contains one symmetry-preserving free parameter. The model is solved
and the Bethe ansatz equations are obtained. 
\end{abstract}




\def\a{\alpha}
\def\b{\beta}
\def\d{\delta}
\def\e{\epsilon}
\def\g{\gamma}
\def\k{\kappa}
\def\l{\lambda}
\def\o{\omega}
\def\t{\theta}
\def\s{\sigma}
\def\D{\Delta}
\def\L{\Lambda}


\def\beq{\begin{equation}}
\def\eeq{\end{equation}}
\def\bea{\begin{eqnarray}}
\def\eea{\end{eqnarray}}
\def\ba{\begin{array}}
\def\ea{\end{array}}
\def\no{\nonumber}
\def\le{\langle}
\def\re{\rangle}
\def\lt{\left}
\def\rt{\right}

\newcommand{\reff}[1]{eq.~(\ref{#1})}

\vskip.3in

Exactly solvable models of strongly correlated fermions have in
recent years generated a great deal of attention, since they
are believed to play a promising role in unraveling the mystery of
high-$T_c$ superconductivity (see, e.g. \cite{book}). 
Several integrable correlated fermion systems have so far appeared in the
literature. Most famous are the supersymmetric $t$-$J$ model and
the Hubbard model. Other integrable
correlated electron systems of interest include the
extended Hubbard model \cite{Ess92}, the supersymmetric $U$ model
proposed in \cite{Bra95} and extensively investigated in 
\cite{Bed95,Ram96,Pfa96}, and its $q$-deformed version \cite{Bar95,Gou96}, 
the model proposed in \cite{Bar91} and its 
generalization \cite{Zhou96}. 

In this communication, we propose an eight state version of the supersymmetric
$U$ model. It is a supersymmetric fermion model with correlated single
particle and pair hoppings as well as uncorrelated triple-particle hopping. 
We then solve the model by means of coordinate Bethe ansatz method and
the Bethe ansatz equations are derived.

Let $c_{j,\a}^\dagger$ ($c_{j,\a}$) denotes a fermionic creation
(annihilation) operator which creates (annihilates) a fermion
of species $\a=+,\;0,\;-$ at
site $j$. These operators satisfy the anti-commutation relations given by
$\{c_{i,\a}^\dagger, c_{j,\b}\}=\d_{ij}\d_{\a\b}$, where 
$i,j=1,2,\cdots,L$ and $\a,\b=+,\;0,\;-$.  At a given lattice site $j$
there are eight possible states:
\bea
&&|0\re\,,~~~c_{j,+}^\dagger|0\re\,,~~~
  c_{j,0}^\dagger|0\re\,,~~~ c_{j,-}^\dagger|0\re,\no\\
&&c_{j,+}^\dagger c_{j,0}^\dagger|0\re\,,~~~ 
  c_{j,+}^\dagger c_{j,-}^\dagger|0\re\,,~~~ 
  c_{j,0}^\dagger c_{j,-}^\dagger|0\re\,,~~~ 
  c_{j,+}^\dagger c_{j,0}^\dagger c_{j,-}^\dagger|0\re\,.\label{states}
\eea
By $n_{j,\a}=c_{i,\a}^\dagger c_{j,\a}$ we denote the number operator
for the fermion of species $\a$ at site $j$.

In the sequel we only consider periodic lattice of length $L$.
The Hamiltonian for our new model reads
\bea
H(g)&=&\sum _{j=1}^L H_{j,j+1}(g),\no\\
H_{j,j+1}(g)&=&-\sum_\a(c_{j,\a}^\dagger c_{j+1,\a}+{\rm h.c.})
  \;exp\lt\{-\frac{\eta}{2} \sum_{\b(\neq\a)}(n_{j,\b}+n_{j+1,\b})
  +\frac{\zeta}{2}
  \sum_{\b\neq\g(\neq\a)}(n_{j,\b}n_{j,\g}
  +n_{j+1,\b} n_{j+1,\g})\rt\}\no\\
& &-\frac{1}{2(g+1)}\sum_{\a\neq\b\neq\g}(c_{j,\a}^\dagger c_{j,\b}^\dagger 
  c_{j+1,\b}c_{j+1,\a}+{\rm h.c.})
  \;exp\lt\{-\frac{\xi}{2}  (n_{j,\g}+n_{j+1,\g})\rt\}\no\\
& &-\frac{2}{(g+1)(g+2)}\lt(c_{j,+}^\dagger c_{j,0}^\dagger 
  c_{j,-}^\dagger c_{j+1,-} c_{j+1,0} c_{j+1,+}+{\rm h.c.}\rt)\no\\
& & +\sum_\a (n_{j,\a}+n_{j+1,\a})-\frac{1}{2(g+1)}\sum_{\a\neq\b}
  (n_{j,\a}n_{j,\b}+n_{j+1,\a}n_{j+1,\b})\no\\
& &+\frac{2}{(g+1)(g+2)}(n_{j,+}n_{j,0}n_{j,-}+n_{j+1,+}n_{j+1,0}
  n_{j+1,-}),\label{hamiltonian}
\eea
where 
\beq
\eta=-\ln\frac{g}{g+1},~~~\zeta=\ln(g+1)-\frac{1}{2}
  \ln g(g+2),~~~\xi=-\ln\frac{g}{g+2}.
\eeq

As will be seen, the supersymmetry algebra underlying this model is
$gl(3|1)$. Remarkably, the model still contains the parameter
$g$ as a free parameter without breaking the supersymmetry. Also this
model is exactly solvable on the one dimensional periodic lattice, as
is seen below.

Our local hamiltonian $H_{i,j}(g)$ does not act as graded
permutation of the states (\ref{states}) at sites $i$ and $j$.
If one projects out any one species, then one singlet, two
doublets and one triplet are projected out and the remaining
four states are: one hole, two singlets and one doublet.
In this case, the projected hamiltonian
is nothing but that of the supersymmetrie $U$ model \cite{Bra95} with
the parameter $U$ (in the supersymmetric $U$ model) related to 
the parameter $g$ here via $U=\pm \frac{2}{g+1}$. It is in this sense that
the present model is an eight state version of the supersymmetric $U$ model. 

To show that (\ref{hamiltonian}) is $gl(3|1)$ supersymmetric, 
we denote the generators of $gl(3|1)$ by $E^\mu_\nu,~~
\mu,\nu=1,2,3,4$ with grading $[1]=[2]=[3]=0,~[4]=1$. In a typical 8-dimensional
representation of $gl(3|1)$, the highest weight $\L=(0,0,0|g)$ itself of the
representation depends on the free parameter $g$, thus giving rise to
a one-parameter family of inequivalent irreps. Let $\{|x\re\}_{x=1}^8$
denote an orthonormal basis with $|1\re, |5\re, |6\re, |7\re$ even
(bosonic) and $|2\re, |3\re, |4\re, |8\re$ odd (fermionic). Then
the simple generators $\{E^\mu_\mu\}_{\mu=1}^4$ and $\{E^\mu_{\mu+1},\;
E^{\mu+1}_\mu\}_{\mu=1}^3$ are $8\times 8$ supermatrices of the form
\bea
&&E^1_2=|3\re \le 4|+|5\re\le 6|,~~~E^2_1=|4\re\le 3|+|6\re\le 5|,~~~
  E^1_1=-|4\re\le 4|-|6\re\le 6|-|7\re\le 7|-|8\re\le 8|,\no\\
&&E^2_3=|2\re\le 3|+|6\re\le 7|,~~~
  E^3_2=|3\re\le 2|+|7\re\le 6|,~~~
  E^2_2=-|3\re\le 3|-|5\re\le 5|-|7\re\le 7|-|8\re\le 8|,\no\\
&&E^3_4=\sqrt{g}\,|1\re\le 2|+\sqrt{g+1}\,(|3\re\le 5|+|4\re\le 6|)
   +\sqrt{g+2}\,|7\re\le 8|,\no\\
&&E^4_3=\sqrt{g}\,|2\re\le 1|+\sqrt{g+1}\,(|5\re\le 3|+|6\re\le 4|)
   +\sqrt{g+2}\,|8\re\le 7|,\no\\
&&E^3_3=-|2\re\le 2|-|5\re\le 5|-|6\re\le 6|-|8\re\le 8|,\no\\
&&E^4_4=g\,|1\re\le 1|+(g+1)\,\left (|2\re\le 2|+|3\re\le 3|
   |4\re\le 4|\right )
  +(g+2)\,(|5\re\le 5|+|6\re\le 6|+|7\re\le 7|)
  +(g+3)\,|8\re\le 8|.\label{matrix}
\eea
The non-simple generators are obtained from the simple ones by using the
defining (anti-)commutation relations of $gl(3|1)$, which we omit. 
Further choose
\bea
&&|1\re=|0\re\,,~~~|2\re=c_{j,+}^\dagger|0\re\,,~~~
  |3\re=c_{j,0}^\dagger|0\re\,,~~~ |4\re=c_{j,-}^\dagger|0\re,\no\\
&&|5\re=c_{j,+}^\dagger c_{j,0}^\dagger|0\re\,,~~~ 
  |6\re=c_{j,+}^\dagger c_{j,-}^\dagger|0\re\,,~~~ 
  |7\re=c_{j,0}^\dagger c_{j,-}^\dagger|0\re\,,~~~ 
  |8\re=c_{j,+}^\dagger c_{j,0}^\dagger c_{j,-}^\dagger|0\re\,.\label{choice}
\eea
Then the verification that the hamiltonian $H(g)$ commutes with all 
generators of $gl(3|1)$ is just a straightforward calculation.

The model is exactly solvable by the Bethe ansatz, since
the local hamiltonian $H_{i,i+1}(g)$
is actually derived from a $gl(3|1)$-invariant rational R-matrix
[which satisfies the graded Yang-Baxter equation]. The R-matrix
is given by \cite{Del95}
\beq
\check{R}(u)=\check{P}_1-\frac{u+2g}{u-2g}\check{P}_2
  +\frac{(u+2g)(u+2g+2)}{(u-2g)(u-2g-2)}\check{P}_3
  -\frac{(u+2g)(u+2g+2)(u+2g+4)}{(u-2g)(u-2g-2)(u-2g-4)}\check{P}_4,
  \label{rational-R}
\eeq
where $\check{P}_k,~k=1,2,3,4$, are four projection operators:
\beq
\check{P}_1=\sum_{k=1}^8 |\Psi_k^1\re\le\Psi_k^1|,~~~
  \check{P}_4=\sum_{k=1}^8|\Psi_k^4\re\le\Psi_k^4|,~~~
  \check{P}_2=\frac{1}{2}(I+P)-\check{P}_1, ~~~
  \check{P}_3=\frac{1}{2}(I-P)-\check{P}_4,\label{projectors}
\eeq
where $P$ is the graded permutation operator and 
$|\Psi_k^1\re,~
  |\Psi_k^4\re,~k=1,2,\cdots,8$ are given by
\bea
&&|\Psi^1_1\re=|1\re\otimes |1\re,~~~~
  |\Psi^1_2\re=\frac{1}{\sqrt{2}}(|2\re\otimes |1\re+|1\re\otimes |2\re),\no\\
&&|\Psi^1_3\re=\frac{1}{\sqrt{2}}(|3\re\otimes |1\re+|1\re\otimes |3\re),~~~~
  |\Psi^1_4\re=\frac{1}{\sqrt{2}}(|4\re\otimes |1\re+|1\re\otimes |4\re),\no\\
&&|\Psi^1_5\re=\frac{1}{\sqrt{2(2g+1)}}[\sqrt{g+1}(|5\re\otimes |1\re
  +|1\re\otimes |5\re)+\sqrt{g}(|2\re\otimes |3\re-|3\re\otimes |2\re)],\no\\
&&|\Psi^1_6\re=\frac{1}{\sqrt{2(2g+1)}}[\sqrt{g+1}(|6\re\otimes |1\re
  +|1\re\otimes |6\re)+\sqrt{g}(|2\re\otimes |4\re-|4\re\otimes |2\re)],\no\\
&&|\Psi^1_7\re=\frac{1}{\sqrt{2(2g+1)}}[\sqrt{g+1}(|7\re\otimes |1\re
  +|1\re\otimes |7\re)+\sqrt{g}(|3\re\otimes |4\re-|4\re\otimes |3\re)],\no\\
&&|\Psi^1_8\re=\frac{1}{2\sqrt{2g+1}}[\sqrt{g}(|2\re\otimes |7\re
  +|7\re\otimes |2\re
  +|5\re\otimes |4\re+|4\re\otimes |5\re
  -|3\re\otimes |6\re-|6\re\otimes |3\re)
  +\sqrt{g+2}(|8\re\otimes |1\re+|1\re\otimes |8\re)],\no\\
&&|\Psi^4_1\re=\frac{1}{2\sqrt{2g+3}}[\sqrt{g+2}(|7\re\otimes |2\re
  -|2\re\otimes |7\re
  +|5\re\otimes |4\re-|4\re\otimes |5\re
  -|6\re\otimes |3\re+|3\re\otimes |6\re)
  +\sqrt{g}(|1\re\otimes |8\re-|8\re\otimes |1\re)],\no\\
&&|\Psi^4_2\re=\frac{1}{\sqrt{2(2g+3)}}[\sqrt{g+1}(|8\re\otimes |2\re
  +|2\re\otimes |8\re)+\sqrt{g+2}(|5\re\otimes |6\re-|6\re\otimes |5\re)],\no\\
&&|\Psi^4_3\re=\frac{1}{\sqrt{2(2g+3)}}[\sqrt{g+1}(|8\re\otimes |3\re
  +|3\re\otimes |8\re)+\sqrt{g+2}(|5\re\otimes |7\re-|7\re\otimes |5\re)],\no\\
&&|\Psi^4_4\re=\frac{1}{\sqrt{2(2g+3)}}[\sqrt{g+1}(|8\re\otimes |4\re
  +|4\re\otimes |8\re)+\sqrt{g+2}(|6\re\otimes |7\re-|7\re\otimes |6\re)],\no\\
&&|\Psi^4_5\re=\frac{1}{\sqrt{2}}(-|8\re\otimes |5\re+|5\re\otimes |8\re),~~~~
  |\Psi^4_6\re=\frac{1}{\sqrt{2}}(-|8\re\otimes |6\re+|6\re\otimes |8\re),\no\\
&&|\Psi^4_7\re=\frac{1}{\sqrt{2}}(-|8\re\otimes |7\re+|7\re\otimes |8\re),~~~~
  |\Psi^4_8\re=|8\re\otimes |8\re,\label{basis}
\eea
which are easily seen to be orthonormal, so that
\bea
&&\le\Psi^1_k|=\left (|\Psi^1_k\re\right )^\dagger,~~~~
  \le\Psi^4_k|=\left (|\Psi^4_k\re\right )^\dagger,~~~k=1,\cdots,8,\no\\
&&\left (|x\re\otimes |y\re\right )^\dagger=(-1)^{[|x\re][|y\re]}
  \le y|\otimes \le x|.\label{dual}
\eea
Here $[|x\re]=0$ for even (bosonic) $|x\re$ and $[|x\re]=1$ for odd
(fermionic) $|x\re$. 

On the $L$-fold tensor product space we denote $\check{R}(u)_{j,j+1}=
I^{\otimes(j-1}\otimes\check{R}(u)\otimes I^{\otimes(L-j-1)}$, and
define the local hamiltonian by
\beq
H^{\rm R}_{j,j+1}(g)=\left .\frac{d}{du}\check{R}_{j,j+1}
   (u)\right |_{u=0}=-4(2g+1)\,(\check{P}_1)_{j,j+1}+\frac{4g(2g+3)}{g+2}
   \,(\check{P}_4)_{j,j+1}+2g\,P_{j,j+1}.\label{h-r}
\eeq
Then by (\ref{projectors}), (\ref{basis}), (\ref{dual}) and (\ref{choice}),
and after tedious but straightforward manipulation,
one gets, up to a constant, $H_{j,j+1}(g)=\frac{1}{2(g+1)}
H^{\rm R}_{j,j+1}(g)$. [This identity also indicates that $H(g)$
commutes with all generators of $gl(3|1)$, since the rational R-matrix
$\check{R}(u)$ is a $gl(3|1)$ invariant.]

As mentioned above, the system is exactly solvable by means of 
Bethe ansatz technique. We assume the following wavefunction
\beq
\psi_{\a_1,\cdots,\a_N}(x_1,\cdots,x_N)=\sum_P\e_P\,A_{\a_{Q_1},\cdots,
  \a_{Q_N}}(k_{P_{Q_1}},\cdots,k_{P_{Q_N}})\,exp\lt(i\sum_{j=1}^N
  k_{P_j}x_j\rt),
\eeq
where $Q$ is the permutation of the $N$ particles such that
$1\leq x_{Q_1}\leq\cdots\leq x_{Q_N}\leq L$. Denote $X_Q=\{x_{Q_1}\leq
\cdots\leq x_{Q_N}\}$. The coefficients $A_{\a_{Q_1},\cdots,\a_{Q_N}}
(k_{P_{Q_1}},\cdots,k_{P_{Q_N}})$ from regions other than $X_Q$ are
connected with each other by elements of two-particle S-matrix:
\beq
S_{1,2}(k_1,k_2)=\frac{\t(k_1)-\t(k_2)+ic{\cal P}_{12}}
  {\t(k_1)-\t(k_2)+ic},
\eeq
where operator ${\cal P}_{12}$ interchanges the species variables
$\a_1$ and $\a_2$ ($\a_1,\;\a_2=+,0,-$), the rapidities $\t(k_j)$
are related to the single-particle quasi-momenta $k_j$ by
$\t(k)=\frac{1}{2}\tan(\frac{k}{2})$ and the dependence on the system
parameter $g$ is incorporated in the parameter $c=e^{\eta}-1=1/g$.
The periodicity condition for the system on the finite interval $(0,L)$
results in the Bethe equations for the set of $N$ momenta $k_j$
: $exp(ik_jL)=T_j,~j=1,\cdots,N$, where
\beq
T_j=S_{j,j+1}(k_j,k_{j+1})\cdots S_{j,N}(k_j,k_N)S_{j,1}(k_j,k_1)\cdots
    S_{j,j-1}(k_j,k_{j-1}), ~~~~j=1,\cdots,N.
\eeq
The meaning of $T_j$ is the scattering matrix of the $j$-th particle on
the other $(N-1)$ particles. So now the problem is to diagonalize $T_j$
to arrive at a system of scalar equations. It is easy to show that
$T_j=\tau(\l=k_j)$, where 
\beq
\tau(\l)=tr_0\lt[S_{0,1}(\l-k_1)\cdots S_{0,N}(\l-k_N)\rt]
\eeq
is the transfer matrix of the inhomogeneous $gl(3)$-spin magnet of $N$ sites.
The commutativity of the transfer matrix for different values of the
spectral parameter $\l$ implies that $T_j,~j=1,\cdots,N$ can be
diagonalized simultaneously.  The
Bethe ansatz equations are written in terms of the rapidities
$\t_j\equiv \t(k_j)$ and $\l_\s$
\bea
e^{ik_j L}&=&\prod_{\s=1}^{M_1}\frac{\t_j-\l^{(1)}_\s+ic/2}
      {\t_j-\l^{(1)}_\s-ic/2},\no\\
\prod_{j=1}^N\frac{\l^{(1)}_\s-\t_j+ic/2}{\l^{(1)}_\s-\t_j-ic/2}&=&
   -\prod_{\rho=1}^{M_1}\frac{\l^{(1)}_\s-\l^{(1)}_\rho+ic}
   {\l^{(1)}_\s-\l^{(1)}_\rho-ic}
   \prod_{\rho=1}^{M_2}\frac{\l^{(1)}_\s-\l^{(2)}_\rho-ic/2}
   {\l^{(1)}_\s-\l^{(2)}_\rho+ic/2},~~~\s=1,\cdots,M_1,\no\\
\prod_{\rho=1}^{M_1}\frac{\l^{(2)}_\g-\l^{(1)}_\rho+ic/2}
   {\l^{(2)}_\g-\l^{(1)}_\rho-ic/2}&=&-
   \prod_{\rho=1}^{M_2}\frac{\l^{(2)}_\g-\l^{(2)}_\rho+ic}
   {\l^{(2)}_\g-\l^{(2)}_\rho-ic},~~~\g=1,\cdots,M_2,\label{Bethe-ansatz}
\eea
The energy of the system in the state corresponding to the sets of
solutions $\{\t_j\}$ and $\{\l_\s\}$ is (up to an additive constant,
which we drop)
$E=-2\sum_{j=1}^N\cos k_j$.

To summarize, we have presented an integrable eight state version of the
supersymmetric $U$ model. It is a $gl(3|1)$ supersymmetric fermion
model with generalized hoppings. We have solved the model by the coordinate
Bethe ansatz method and derived the Bethe ansatz equations. 
There are many things remained to be done for this new model. One of
them is to study physical properties, such as the phase diagram and
the critical exponents, of the system. It is also interesting to
incorporate integrable boundary conditions into the model and to investigate
finite-size corrections of the boundary system. We hope to report
results on those aspects in future publications.

\vskip.3in
This work is supported by Australian Research Council, University of
Queensland New Staff Research Grant and External Support Enabling Grant.
H.-Q.Z is also supported by China National NSF and 
would like to thank Department of Mathematics of UQ for kind hospitality.


\end{document}